\documentclass[11pt]{article}

\usepackage{setspace}
\usepackage{mathrsfs}
\usepackage{amssymb}
\usepackage{color}
\usepackage{amsmath}
\usepackage{ascmac}
\usepackage{amsthm}
\usepackage[dvips]{graphicx}
\usepackage{booktabs}


\def\bpsi{{\text{\boldmath $\psi$}}}
\def\bpsih{\widehat{\bpsi}}
\def\T{{\text{\boldmath $T$}}}
\def\W{{\text{\boldmath $W$}}}
\def\t{{\text{\boldmath $t$}}}
\def\x{{\text{\boldmath $x$}}}
\def\bbe{{\text{\boldmath $\beta$}}}
\def\la{\lambda}

\title{A flexible family of distributions on the cylinder}

\author{Shonosuke Sugasawa\thanks{Graduate School of Economics, The University of Tokyo, E-mail: shonosuke622@gmail.com} , \  Kunio Shimizu\thanks{The Institute of Statistical Mathematics} \ and Shogo Kato$^{\dagger}$
}

\date{}

\begin{document}

\maketitle

\begin{abstract}
We propose a flexible family of distributions, generalized $t$-distributions, on the cylinder which is obtained as a conditional distribution of a trivariate $t$ distribution. 
The new distribution has unimodality or bimodality, symmetry or asymmetry, depending on the values of parameters and flexibly fits the cylindrical data. 
The circular marginal of this distribution is distributed as a generalized $t$-distribution on the circle. Some other properties are also investigated. The proposed distribution is applied to the real cylindrical data.

\par\vspace{4mm}
\end{abstract}

\section{Introduction}\label{Introduction}
Directional or circular data often appear in a variety of scientific fields and various stochastic models have been proposed for analyzing such data. 
For univariate circular data, there have been many distributions investigated in terms of both tractability and applicability, see Jones and Pewsey (2005), Kato and Jones (2010) and Kato and Jones (2015).
However, we sometimes encounter situations which involve both circular and linear variables, namely cylindrical data such as the pair of wind direction and temperature (Mardia and Sutton, 1978) or directions and distances of animal movements (Fisher, 1993). 
For such data, the distribution on the cylinder is needed, but there are not so many distributions compared to univariate circular distributions.
We give a brief review below for several cylindrical distributions known in the literature. 
Johnson and Wehrly (1978) gave a distribution based on the principle of maximum entropy subject to constraints on certain moments, and Mardia and Sutton (1978) provided another distribution as a conditional distribution of a trivariate normal distribution or a maximum entropy distribution. An extension of the distribution by Mardia and Sutton (1978) was studied by Kato and Shimizu (2008), which can also be derived as a maximum entropy distribution or a conditional of a trivariate normal.

In this paper, we propose the generalized $t$-distribution on the cylinder, which is a natural extension of the member of the exponential family given by Kato and Shimizu (2008). 
The proposed distribution is also regarded as a cylindrical extension of the generalized $t$-distribution on the circle proposed by Siew et al. (2008).
In fact, the circular marginal distribution is the generalized $t$-distribution on the circle.   
The proposed distribution can be obtained as a conditional distribution of a trivariate $t$ distribution and characterized as the maximum $\beta$-entropy distribution.
This is a quite flexible distribution which allows for both asymmetry and variations in tail weight in terms of parameter values. 
We investigate some properties such as marginal and conditional distributions, modality, moments, circular-linear correlation and skewness.  
We briefly discuss a circular-linear regression model derived from the conditional distribution.
For practical use, we provide an iterative algorithm for parameter estimation.

Subsequent sections are organized as follows. Section 2 provides a derivation of the distribution. 
In Section 3 some properties of the new distribution are studied. 
We apply the proposed distribution to the data set of the wind direction and ozone level given in Johnson and Wehrly (1977) in Section 4.

\section{Derivation}
\label{Derivation}
Suppose that a trivariate random vector $W$ is distributed as a trivariate $t$ distribution with degrees of freedom $\alpha$, mean vector $\eta = (\eta_1,\eta_2,\eta_3)' \in \mathbb{R}^3$ and variance-covariance matrix $\Sigma$, where
$$
\Sigma =\left( \begin{array}{ccc}
\sigma_1^2 & \rho_{12} \sigma_1 \sigma_2 & \rho_{13} \sigma_1 \sigma_3 \\
\rho_{12} \sigma_1 \sigma_2 & \sigma_2^2 & \rho_{23} \sigma_2 \sigma_3 \\
\rho_{13} \sigma_1 \sigma_3 & \rho_{23} \sigma_2 \sigma_3 & \sigma_3^2
\end{array} \right) 
$$
for $\sigma_j>0$ ($j=1,2,3$), $-1<\rho_{12}<1$ and $1+2\rho_{12}\rho_{13}\rho_{23}-\rho_{12}^2-\rho_{13}^2-\rho_{23}^2>0.$
From Kotz and Nadarajah (2004), the distribution of $W$ has density
$$
f(w)=\frac{\Gamma((\alpha+3)/2)}{(\alpha\pi)^{3/2}|\Sigma|^{1/2}\Gamma(\alpha/2)}\left\{ 1+\frac{(w-\eta)'\Sigma^{-1}(w-\eta)}{\alpha} \right\}^{-(\alpha+3/2)} , \quad w \in \mathbb{R}^3.
$$

We use a cylindrical coordinate $W=(X,X_1,X_2)'$, where $X_1=R \cos \Theta$ and $X_2=R \sin \Theta$ with $R>0$ and $0 \le \Theta <2\pi$, and consider the conditional distribution of $(X,\Theta)'$ given $R=r$, which provides a distribution on the cylinder.
Define new parameters as
\begin{center}
$
\begin{array}{l}
\mu(\theta) = \mu +\lambda \cos (\theta-\nu), \quad
\tau^2 = \displaystyle{\frac{\sigma_1^2 \rho^2}{1-\rho_{23}^2}} , \\
\kappa_1^{\ast} \cos \mu_1 =r(b_1\eta_2-b_2\eta_3), \quad
\kappa_1^{\ast} \sin \mu_1 =r(b_3\eta_3-b_2\eta_2), \\
\displaystyle{
\kappa_2^{\ast} \cos 2\mu_2 = \frac{r^2 (b_3-b_1)}{4}, \quad
\kappa_2^{\ast} \sin 2\mu_2 = \frac{r^2 b_2}{2} }
\end{array}
$
\end{center}
with 
{$-\infty < \mu < \infty$, $\lambda, \kappa_1^{\ast} ,\kappa_2^{\ast}\ge 0$, $0 \le \nu, \mu_1 <2\pi$}
and $0 \le \mu_2 <\pi$, where
\begin{center}
$
\begin{array}{l}
\mu = \eta_1+a_1 \eta_2 +a_2 \eta_3, \quad
\rho^2 = 1+2 \rho_{12}\rho_{13}\rho_{23}-\rho_{12}^2-\rho_{13}^2-\rho_{23}^2 , \\
\displaystyle{
a_1=\frac{\sigma_1}{\sigma_2}\frac{\rho_{13}\rho_{23}-\rho_{12}}{1-\rho_{23}^2}, \quad
a_2=\frac{\sigma_1}{\sigma_3}\frac{\rho_{12}\rho_{23}-\rho_{13}}{1-\rho_{23}^2},
} \\
\displaystyle{
b_1=\frac{1}{\sigma_2^2(1-\rho_{23}^2)}, \quad
b_2=\frac{\rho_{23}}{\sigma_2 \sigma_3(1-\rho_{23}^2)}, \quad
b_3=\frac{1}{\sigma_3^2(1-\rho_{23}^2)},
} \\
\lambda \cos \nu =-a_1 r, \quad \lambda \sin \nu =-a_2 r.
\end{array}
$
\end{center}
Then we have
$$
\frac12 (w-\eta)'\Sigma^{-1}(w-\eta)
= \frac{1}{2\tau^2} \{ x-\mu(\theta) \}^2 
-\kappa_1^{\ast} \cos (\theta-\mu_1)
-\kappa_2^{\ast} \cos 2(\theta-\mu_2)
+d,
$$
where $d=(b_1\eta_2^2+b_3\eta_3^2-2b_2\eta_2\eta_3)/2 \ (\ge 0)$. The conditional probability density function $f(x,\theta|r)$ of $(X,\Theta)'|(R=r)$ is represented as 
\begin{align}
\label{density}
f(x,&\theta|r)\notag\\
&=C^{-1}
\left[1+
\frac{1}{2\sigma^2} \left\{ x-\mu(\theta) \right \}^2 
-\kappa_1\cos (\theta-\mu_1)-\kappa_2\cos 2(\theta-\mu_2)
 \right]^{-(\alpha+3)/2},  
\end{align}
where $\kappa_1=\kappa_1^{\ast}/\gamma, \ \kappa_2=\kappa_2^{\ast}/\gamma, \ \sigma^2=\gamma\tau^2, \ \gamma=\alpha+d \ (>0)$ and the normalizing constant is 
\begin{eqnarray}
\label{NC}
C&=& 2\sqrt{2}\pi B(1/2,\alpha/2+1)\sigma 
\left\{
F_4 \left( \frac{\alpha}{4}+\frac12, \frac{\alpha}{4}+1, 1,1; \kappa_1^2,\kappa_2^2 \right) \right. \nonumber \\
&&
+2 \sum_{j=1}^{\infty} \frac{(\alpha/2+1)_{3j}}{(2j)!j!}
\left( \frac{\kappa_1}{2} \right)^{2j} 
\left( \frac{\kappa_2}{2} \right)^{j} \cos 2j (\mu_2-\mu_1) \nonumber \\
&& \times \left.
F_4 \left( \frac{\alpha+6j}{4}+\frac12, \frac{\alpha+6j}{4}+1, 2j+1,j+1;  \kappa_1^2,\kappa_2^2 \right) \right\}.
\end{eqnarray}
Here $F_4$ denotes Appell's double hypergeometric function (Gradshteyn and Ryzhik, 2007, 9.180.4)  defined by
$$
F_4(\alpha_1,\alpha_2,\beta_1,\beta_2;z_1,z_2)=\sum_{i=0}^{\infty}\sum_{j=0}^{\infty}
\frac{(\alpha_1)_{i+j}(\alpha_2)_{i+j}}{(\beta_1)_{i}(\beta_2)_{j}} \frac{z_1^i z_2^j}{i!j!}, \quad
\sqrt{z_1}+\sqrt{z_2}<1
$$
with Pochhammer's symbol
$$
(c)_j = \left\{ \begin{array}{ll} c(c+1)\cdots (c+j-1), & j \ge 1, \\ 1,& j=0, \end{array} \right. 
$$
and $B$ the beta function.
The distribution with density function (\ref{density}) has nine parameters
{$\alpha ,\sigma >0$, $-\infty < \mu < \infty$, $\kappa_1,\kappa_2 ,\lambda \ge 0$, $0 \le \nu, \mu_1 <2\pi$}
and $0 \le \mu_2 <\pi$ with restriction $\kappa_1+\kappa_2<1$.
The resulting distribution should be called the generalized $t$-distribution on the cylinder.
Relationships between the new and original parameters are:
\begin{itemize}
\item[(a)]
$\eta_2=\eta_3=0 \Leftrightarrow  \kappa_1=0$.

\item[(b)]
$\kappa_2 =0 \Leftrightarrow \rho_{23}= 0, \sigma_2 = \sigma_3$.

\item[(c)]
$\lambda =0 \Leftrightarrow \rho_{12}=\rho_{13}=0$.
\end{itemize}

As being introduced in (\ref{density}), the parameter $\alpha$ was assumed positive. 
However, the density is still valid when the parameter space of $\alpha$ is extended to $\alpha\geq-1$. 
Note that $\alpha$ corresponds to the degrees of freedom in the generalized $t$-distribution on the cylinder (\ref{density}) and the parameter $\alpha$ determines the degree of concentration around the mode. 
The proposed distribution with density (\ref{density}) has 9 parameters.
The roles of the parameters are as follows: $(\mu_1,\mu_2)$ and $\mu$ are location parameters of $\Theta$ and $X$, respectively, and $\sigma$ is the scale parameter of $X$. 
$\la$ and $\nu$ in $\mu(\theta)$ determines the degree of correlation between $X$ and $\Theta$, and $\lambda$ controls the skewness of $X$.
$\kappa_1$ and $\kappa_2$ determine the morality of the distribution, discussed in Section 3.3.
$\alpha$ controls the tail weight of the distribution.
To see the role of $\alpha$ and $\lambda$, we provide contour plots of the proposed density (\ref{density}) in Figure 1 for some combinations of $\alpha$ and $\lambda$. 
Other parameters are specified as $\mu=\nu=\mu_1=\mu_2=0, \ \sigma^2=1, \  \kappa_1=0.1, \ \kappa_2=0.4$.
The column (I) in Figure 1 illustrates the interpretation of $\alpha$: as $\alpha$ increases, the tail weight of the density increases. 
The role of $\lambda$ as controlling the skewness of the distribution can be seen from the column (II) in Figure 1. 
The interpretation of $\kappa_1,\kappa_2$ are discussed in Section 3.3.

Under the assumption $\gamma = (\alpha +3)/2$, if we use reparametrization $1/\psi =-(\alpha+3)/2, \ -\kappa_1=\tanh (\kappa_1^{\ast \ast}\psi), \ -\kappa_2=\tanh (\kappa_2^{\ast \ast}\psi)$, then the density (\ref{density}) has a similar expression to the family of distributions introduced by Jones and Pewsey (2005):
$$
f(x,\theta |r) \propto \left[ 1-\frac{\psi}{2\tau^2} \{ x-\mu(\theta) \}^2+\tanh (\kappa_1^{\ast \ast}\psi) \cos (\theta-\mu_1)+\tanh (\kappa_2^{\ast \ast}\psi) \cos 2(\theta-\mu_2) \right]^{1/\psi}.
$$
The restriction $\alpha \ge -1$ in (\ref{density}) is changed into $-1\le \psi <0$.

\begin{figure}[!thb]
\centering

(I)\\
\includegraphics[width=5.3cm,clip]{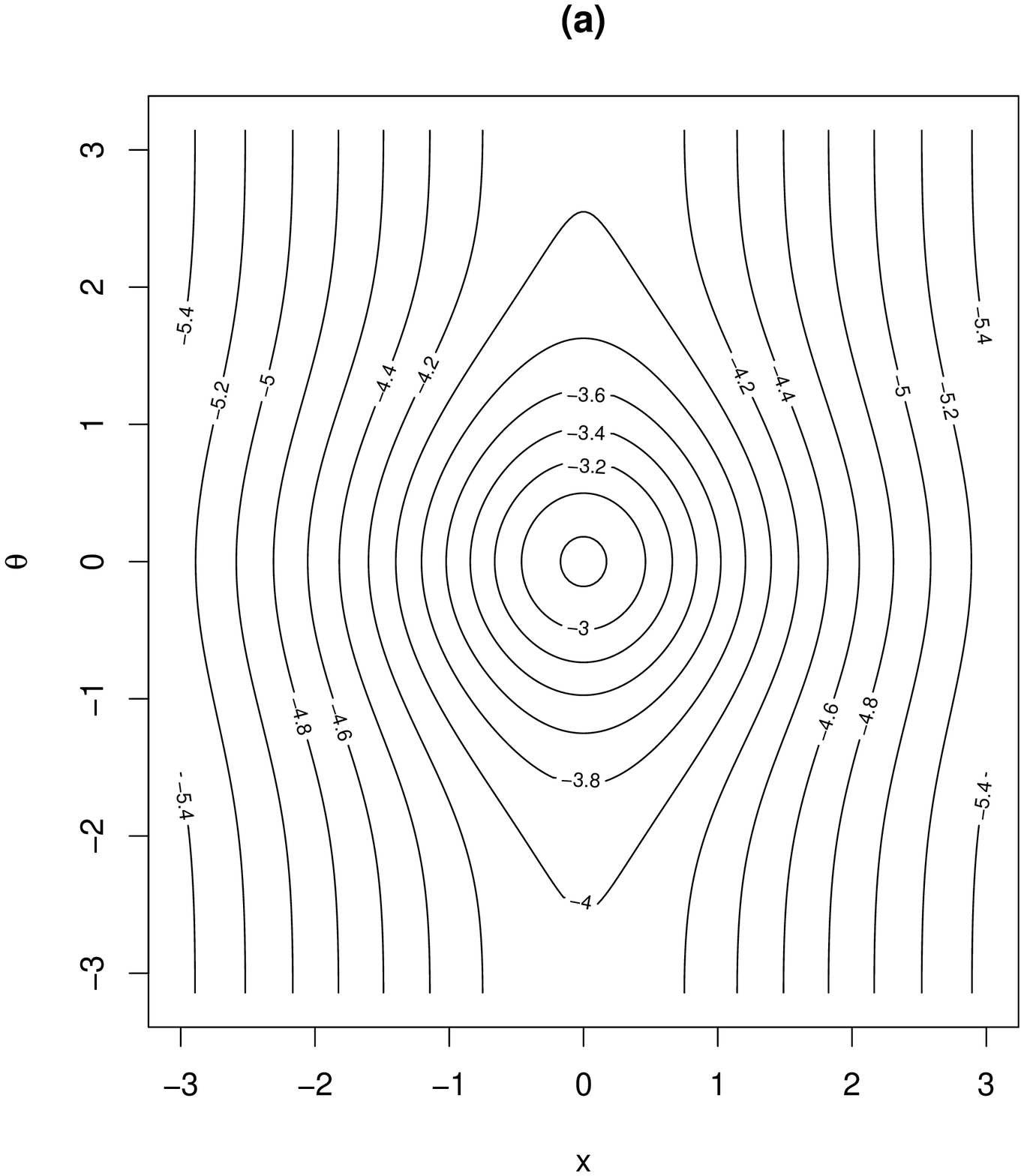}
\includegraphics[width=5.3cm,clip]{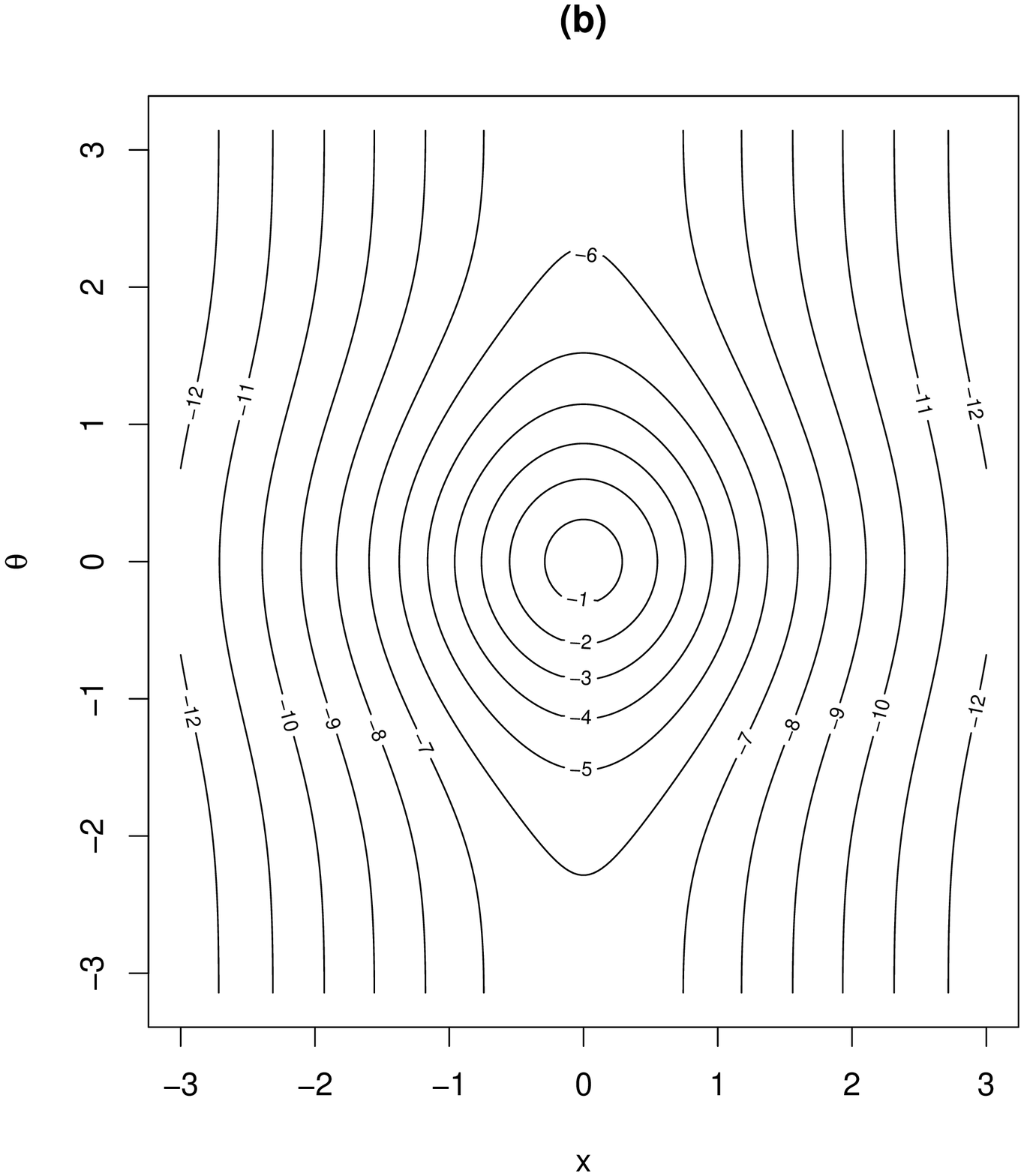}
\includegraphics[width=5.3cm,clip]{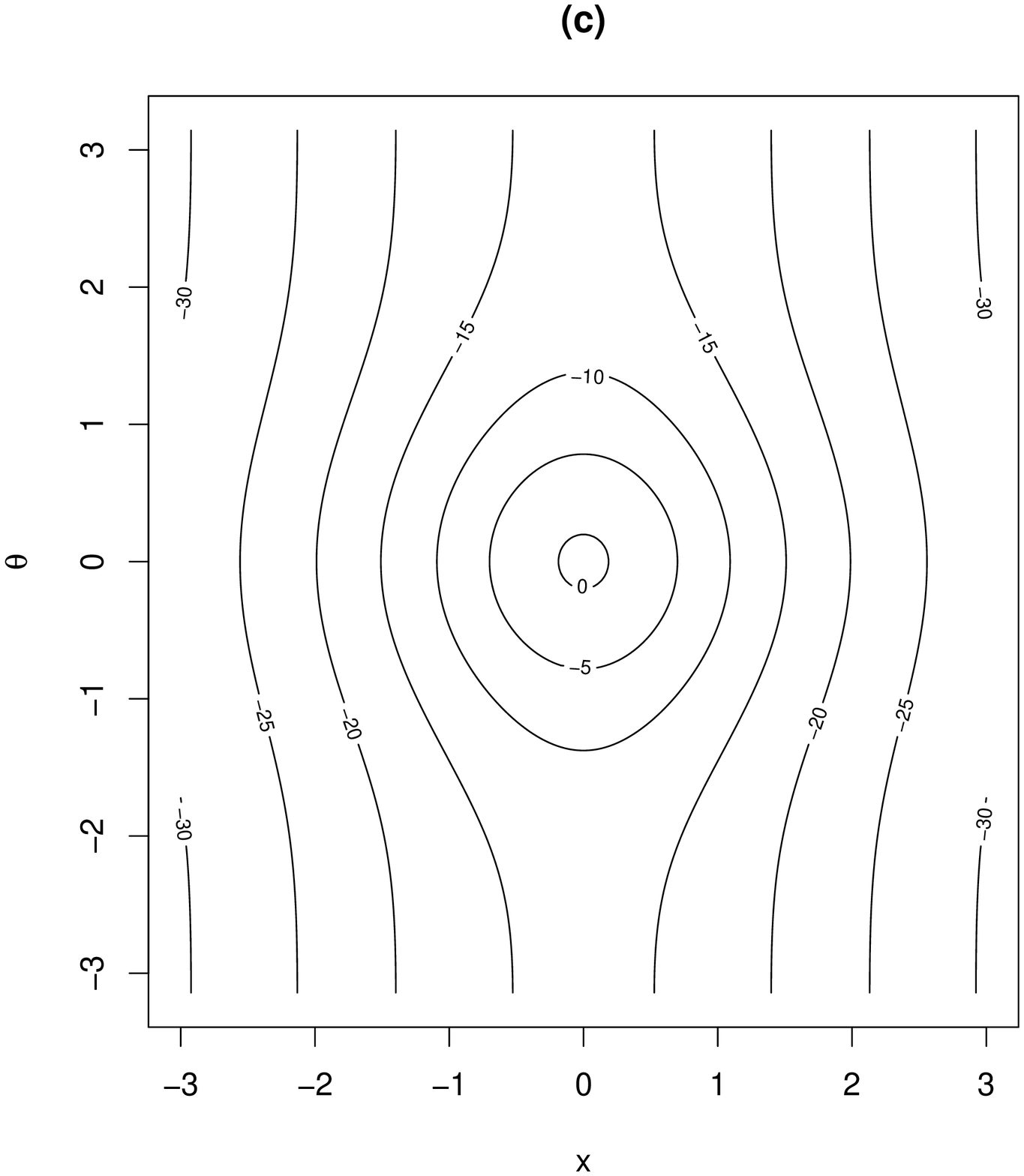}\\
(II)\\
\includegraphics[width=5.3cm,clip]{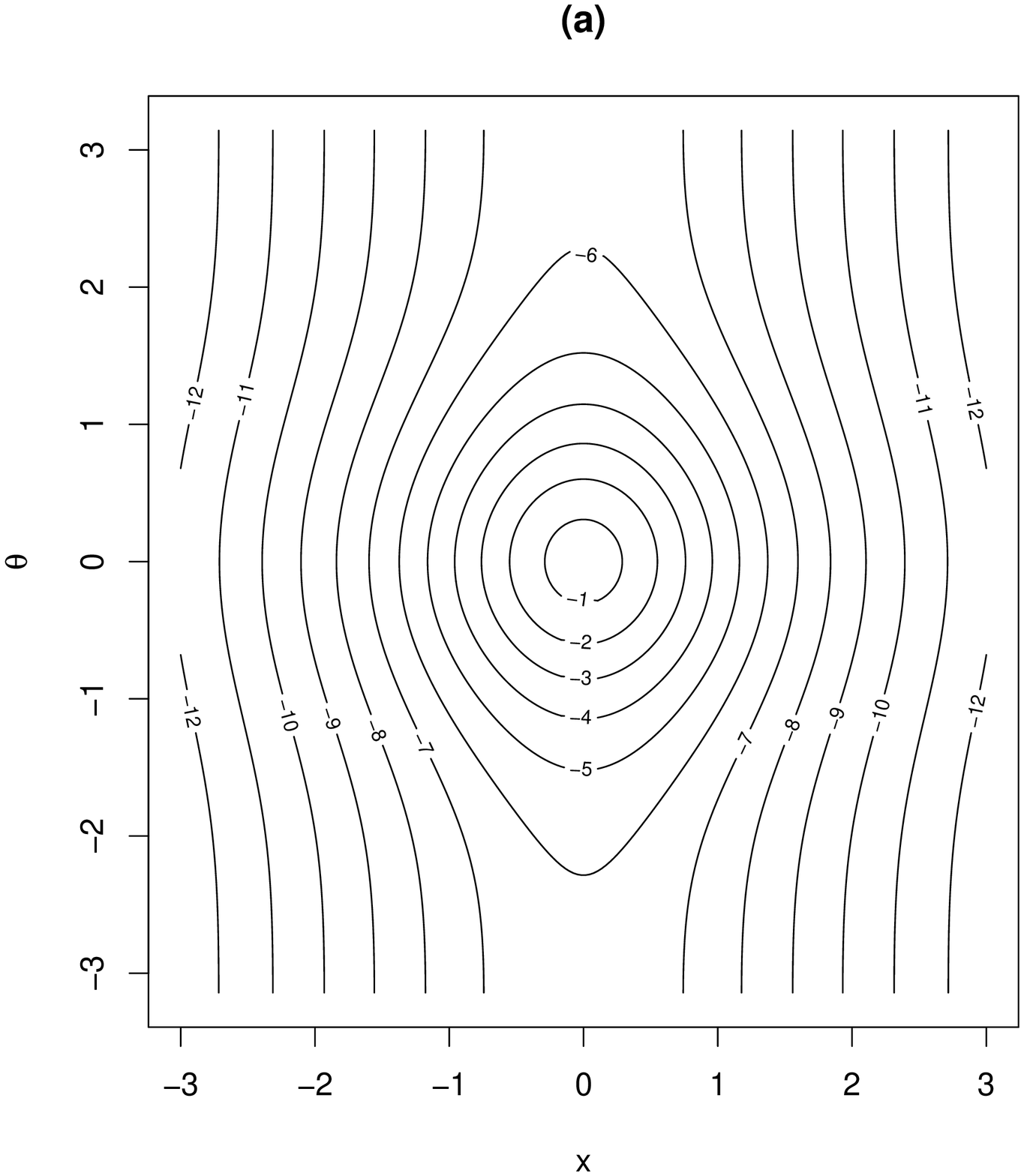}
\includegraphics[width=5.3cm,clip]{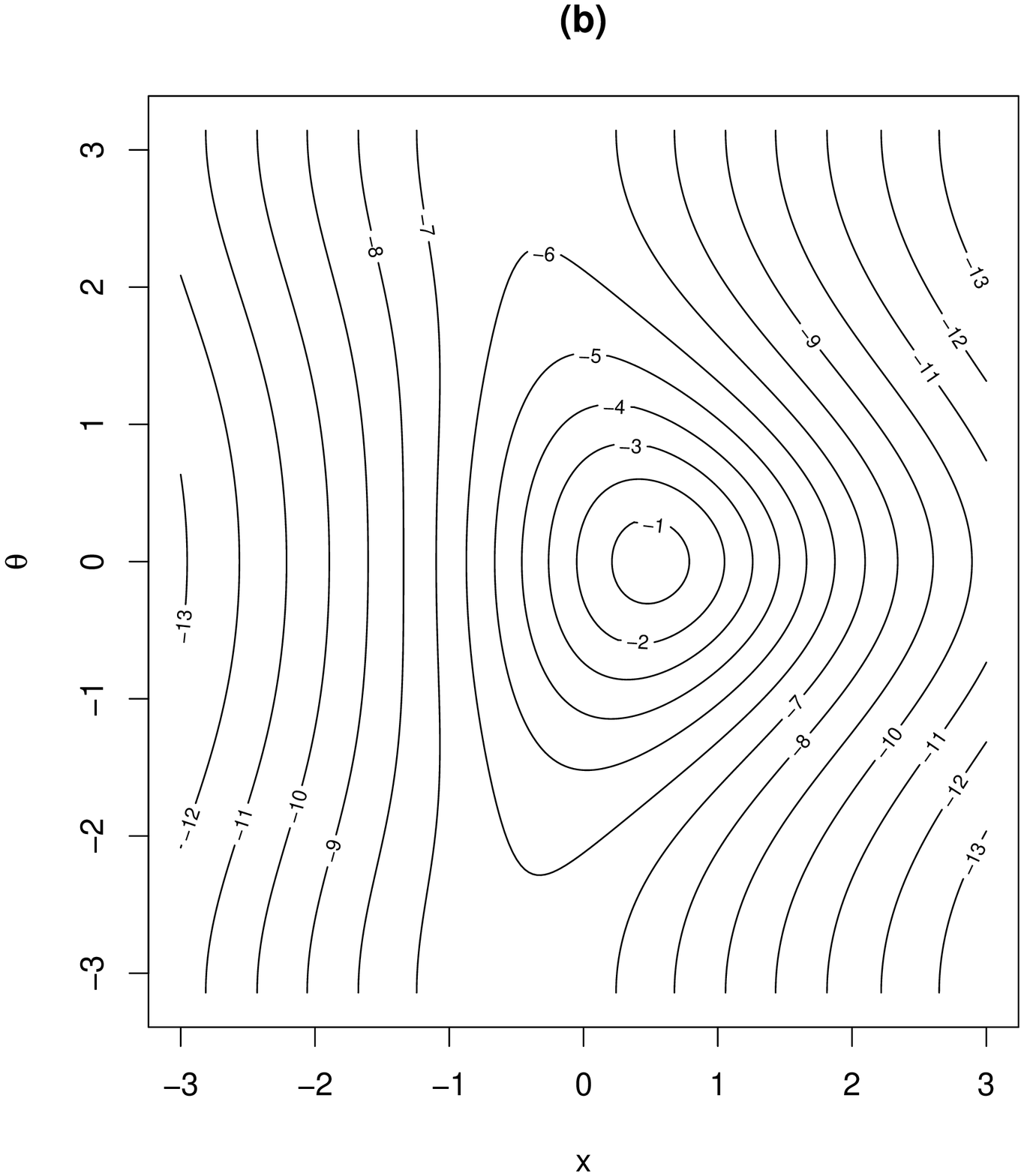}
\includegraphics[width=5.3cm,clip]{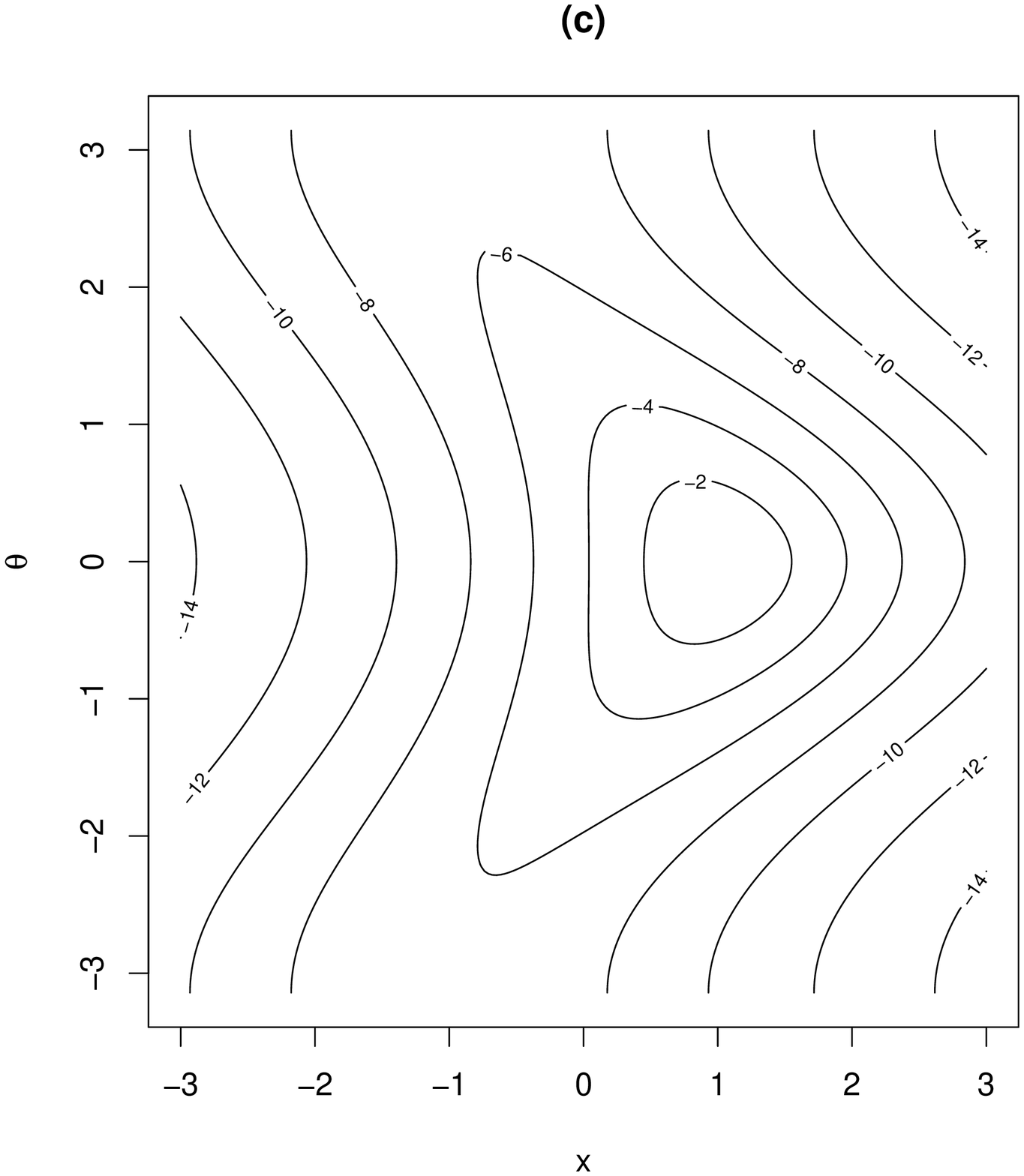}
\caption{Contour plots of density (\ref{density}) for: (I) $\la=0$ and (a) $\alpha=-1$, (b) $\alpha=10$, and (c) $\alpha=20$; (II) $\alpha=6$ and (a) $\la=0$, (b) $\la=0.5$, and (c) $\la=1$.
The other parameters are set as $\mu=\nu=\mu_1=\mu_2=0, \ \sigma^2=1, \  \kappa_1=0.1, \ \kappa_2=0.4$.}
\end{figure}

\section{Properties}
\label{Properties}

\subsection{Special cases}
\noindent
1. Under the assumption $\gamma=\alpha/2$ in (\ref{density}), letting $\gamma \ (=\alpha/2)\to\infty$, we have an extension of the distribution by Mardia and Sutton (1978). Its joint probability density function  (Kato and Shimizu, 2008) is
\begin{equation}
\label{KS-dist}
f(x,\theta)=C_1^{-1} \exp \left[ -\frac{\{x-\mu(\theta)\}^2}{2\tau^2}+\kappa_1^{\ast} \cos (\theta-\mu_1)+\kappa_2^{\ast} \cos 2(\theta-\mu_2) \right]
\end{equation}
with the normalizing constant
\begin{equation}
\label{SC1NC}
C_1=(2\pi)^{3/2}\tau \left[ I_0(\kappa_1^{\ast})I_0(\kappa_2^{\ast})+2\sum_{j=1}^{\infty}
I_j(\kappa_2^{\ast})I_{2j}(\kappa_1^{\ast})\cos2j(\mu_1-\mu_2)\right].
\end{equation}
Here $I_j$ denotes the modified Bessel function of the first kind and order $j$ given by
$$
I_j(z)=\frac{1}{2\pi} \int_0^{2\pi}\cos (j\theta){\rm e}^{z \cos \theta}{\rm d}\theta
=\sum_{r=0}^{\infty}\frac{1}{\Gamma (r+j+1)r!} \left( \frac{z}{2} \right)^{2r+j}, \quad z \in \mathbb{C}.
$$
2. When $\kappa_2=0$,  (\ref{density}) reduces to
\begin{equation}
\label{SC2}
f(x,\theta)=C_2^{-1}
\left[ 1+ 
\frac{1}{2\sigma^2} \left\{ x-\mu(\theta) \right \}^2 
-\kappa_1\cos (\theta-\mu_1)
 \right]^{-(\alpha+3)/2}.
\end{equation}
The normalizing constant is represented as
$$
C_2=2\sqrt{2}\pi\sigma B(1/2,\alpha/2+1)  {}_2F_1 \left( \frac{\alpha}{4}+\frac12,  \frac{\alpha}{4}+1,1;\kappa_1^2 \right)
$$
using the Gauss hypergeometric function ${}_2F_1$.
If we replace $\kappa_1=\kappa_1^{\ast}/\gamma,\sigma^2=\gamma\tau^2$ and let $\gamma=\alpha/2\to\infty$, we have the distribution proposed by Mardia and Sutton (1978) and the constant $C_2$ tends to $(2\pi)^{3/2}\tau I_0(\kappa_1^{\ast})$. This agrees with the fact that
$C_1=(2\pi)^{3/2}\tau I_0(\kappa_1^{\ast})$ when $\kappa_2^{\ast}=0$ in (\ref{SC1NC}).

\subsection{Marginal and conditional distributions}
The marginal distribution of $\Theta$ is 
\begin{equation}
\label{marginal}
f_{\Theta}(\theta)=C_{\Theta}^{-1}\{1-\kappa_1\cos(\theta-\mu_1)-\kappa_2\cos2(\theta-\mu_2)\}^{-\alpha/2-1},
\end{equation}
where
\begin{align}
\label{marNC}
C_{\Theta}&=2\pi
\left\{
F_4 \left( \frac{\alpha}{4}+\frac12, \frac{\alpha}{4}+1, 1,1; \kappa_1^2,\kappa_2^2 \right) \right. \nonumber \\
&
+2 \sum_{j=1}^{\infty} \frac{(\alpha/2+1)_{3j}}{(2j)!j!}
\left( \frac{\kappa_1}{2} \right)^{2j} 
\left( \frac{\kappa_2}{2} \right)^{j} \cos 2j (\mu_2-\mu_1) \nonumber \\
& \times \left.
F_4 \left( \frac{\alpha+6j}{4}+\frac12, \frac{\alpha+6j}{4}+1, 2j+1,j+1;  \kappa_1^2,\kappa_2^2 \right) \right\}.
\end{align}
The distribution with density (\ref{marginal}) is a member of the generalized $t$-distributions on the circle proposed by Siew et al. (2008), and is possibly bimodal and asymmetric. Cosine and sine moments of the generalized $t$-distributions are given in their paper. 
The generalized {\it t}-distributions include the generalized von Mises distribution (cf. Yfantis and Borgman, 1982) as a special case. As another special case when $\kappa_2=0$ in (\ref{density}), the marginal distribution of $\Theta$ belongs to the family of symmetric distributions by Jones and Pewsey (2005). 
Note that the marginal density (\ref{marginal}) is independent of $\lambda, \mu, \sigma$ and $\nu$ which are the parameters of the proposed density (\ref{density}).
On the other hand, the marginal distribution of $X$ does not have a closed form in general. When $\lambda=0$, we can obtain the marginal distribution of $X$ in a closed form given by
$$
f_X(x)=C^{-1}D(x)\left\{1+\frac{1}{2\sigma^2}(x-\mu)^2\right\}^{-(\alpha+3/2)},
$$
where $C$ is defined by (\ref{NC}) and $D(x)$ is obtained by replacing $\alpha$ and $\kappa_i \ (i=1,2)$ with $\alpha+1/2$ and $\kappa_i/\{1+(x-\mu)^2/(2\sigma^2) \}$, respectively, in $C_{\Theta}$ defined in (\ref{marNC}). Note that this density is symmetric about $\mu$.

In (\ref{density}), the conditional distribution of $X$ given $\Theta=\theta$ has the generalized $t$-density function provided by
\begin{equation}
f_{X|\Theta}(x|\theta)=C_{X|\Theta}^{-1}
\left( 1+ \frac{ \{x-\mu(\theta) \}^2}{2 \sigma^2 \left[ 1-  \left\{ \kappa_1 \cos (\theta-\mu_1) + \kappa_2 \cos 2 (\theta-\mu_2) \right\} \right] } \right)^{-(\alpha+3)/2}  \label{cond}
\end{equation}
with the normalizing constant
$$
C_{X|\Theta}=\sqrt{2}\sigma B \left( \frac12,\frac{\alpha}{2}+1 \right) \left[ 1-\{ \kappa_1 \cos (\theta-\mu_1) + \kappa_2 \cos 2 (\theta-\mu_2) \} \right]^{1/2}.
$$
The conditional distribution of $\Theta$ given $X=x$ is a member of the generalized $t$-distributions on the circle given by
$$
f(\theta|X=x)\propto \{1-\kappa_1(x)\cos(\theta-\mu_1)-\kappa_2(x)\cos2(\theta-\mu_2)\}^{-(\alpha+3)/2},
$$
where $\kappa_i(x)=\kappa_i/\{1+(x-\mu)^2/(2\sigma^2) \}, \ i=1,2$.
As shown in Siew et al. (2008), the mean direction of $\Theta$ with density (\ref{marginal}) depends on $\kappa_i, \ i=1,2$.
Thus the conditional mean direction $E(\Theta|X=x)$ depends on $x$ through $\kappa_i(x)$.
The result that the conditional distributions of $X|\Theta=\theta$ and $\Theta|X=x$ are the generalized $t$-distribution comes from the fact that the conditional distribution of a multivariate $t$ distribution is again a multivariate $t$ distribution (Joe, 2015).

\subsection{Modality}
We consider the modality of the distribution with density (\ref{density}). The mode $x^{\ast}$ of (\ref{density}), whenever the value of $\theta$ is specified, is $x^{\ast}=\mu(\theta)$. Similar to Siew et al. (2008), we discuss maximization of the function $m(\theta)=\kappa_1\cos(\theta-\mu_1)+\kappa_2\cos2(\theta-\mu_2)$ with respect to $\theta$. The solution $\theta^{\ast}$ of an equation
\begin{equation}
\kappa_1\sin(\theta-\mu_1)+2\kappa_2\sin2(\theta-\mu_2)=0  \label{mode}
\end{equation}
is a value which maximizes $m(\theta)$ if the sign of $h(\theta^{\ast})$ is positive, where
$$
h(\theta)=\kappa_1\cos(\theta-\mu_1)+4\kappa_2\cos2(\theta-\mu_2).
$$
Equation (\ref{mode}) can be solved numerically for any combinations of $\mu_1$ and $\mu_2$. Without loss of generality, we let $\mu_1=0$. Then (\ref{mode}) has closed form solutions when $\mu_2=0,\pi/2,\pi/4$ and $3\pi/4$. The results are given in Table 1, where $\theta_0=\arccos \{ \kappa_1/(4\kappa_2) \}$, $\theta_1=\arcsin\{(-\kappa_1+\sqrt{\kappa_1^2+32\kappa_2^2})/(8\kappa_2)\}$ and  $\theta_2=\arcsin\{(\kappa_1+\sqrt{\kappa_1^2+32\kappa_2^2})/(8\kappa_2)\}$.
See Yfantis {and} Borgman (1982) for more discussion as to the
{solutions of (\ref{mode}).}

\begin{table}[top]
\caption{ Summary of the modality of the proposed distribution for some values of $\mu_2$ when $\mu_1=0$.}
\begin{center}
\begin{tabular}{c|c|c}\toprule
$\mu_2\ \  $  &  Condition &Modes \ $(x,\theta)$\\  \midrule
$0$& $4\kappa_2>\kappa_1$  &  $(\mu+\lambda\cos\nu,0) , (\mu-\lambda\cos\nu,\pi)  $\\
        &  $4\kappa_2<\kappa_1$  &   $(\mu+\lambda\cos\nu,0)$  \\  \midrule
$\pi/2$& $4\kappa_2>\kappa_1$  &  $(\mu+\lambda\cos(\theta_0-\nu),\theta_0), (\mu+\lambda\cos(\theta_0+\nu),2\pi-\theta_0)$\\
        &  $4\kappa_2<\kappa_1$  &   $(\mu+\lambda\cos\nu,0)$  \\  \midrule
$\pi/4$& $2\kappa_2>\kappa_1$  &  $(\mu+\lambda\cos(\theta_1-\nu),\theta_1),(\mu-\lambda\cos(\theta_2-\nu),\pi+\theta_2)$\\
        &  $2\kappa_2<\kappa_1$  &   $(\mu+\lambda\cos(\theta_1-\nu),\theta_1)$  \\  \midrule
$3\pi/4$& $2\kappa_2>\kappa_1$  &  $(\mu-\lambda\cos(\theta_2+\nu),\pi-\theta_2),(\mu+\lambda\cos(\theta_1+\nu),2\pi-\theta_1)$\\
        &  $2\kappa_2<\kappa_1$  &   $(\mu+\lambda\cos(\theta_1+\nu),2\pi-\theta_1)$  \\   \bottomrule
\end{tabular}
\end{center}
\end{table}

Contour plots and marginal density plots of the proposed distribution are given in Figure 2. 
Figure 2 shows the case when $\kappa_1=0.5$, $\kappa_2=0.1$, $\lambda=0$, and (II) $\kappa_1=0.2$, $\kappa_2=0.3$, $\lambda=1$, while the other parameters are set as $\sigma=1$,  $\mu=0$, $\nu=\pi/3$, $\mu_1=\mu_2=0$, $\alpha=6$.
In the case of (I) in Figure 2, the joint density is unimodal and the marginal densities of $X$ and $\Theta$ are symmetric. 
On the other hand, in the case of (II) in Figure 3, the joint density is bimodal and the marginal density of $X$ is asymmetric. 
In fact, the marginal density of $X$ is possibly skew depending on the values of the parameters, as will be seen in Section 3.5. The marginal density of $\Theta$ in Figure 3 is bimodal as given in Table 1.

In the case where a unimodal distribution is desired, we put, for example, $\mu_2=\mu_1+\pi/4$ with restriction $2|\kappa_2|<\kappa_1$ to get a unimodal density function from (\ref{density}).
This submodel can be useful for modeling multimodality when a finite mixture model is easier to interpret than a bimodal distribution.

\begin{figure}[!thb]
\centering
(I)\\
\includegraphics[width=5cm,clip]{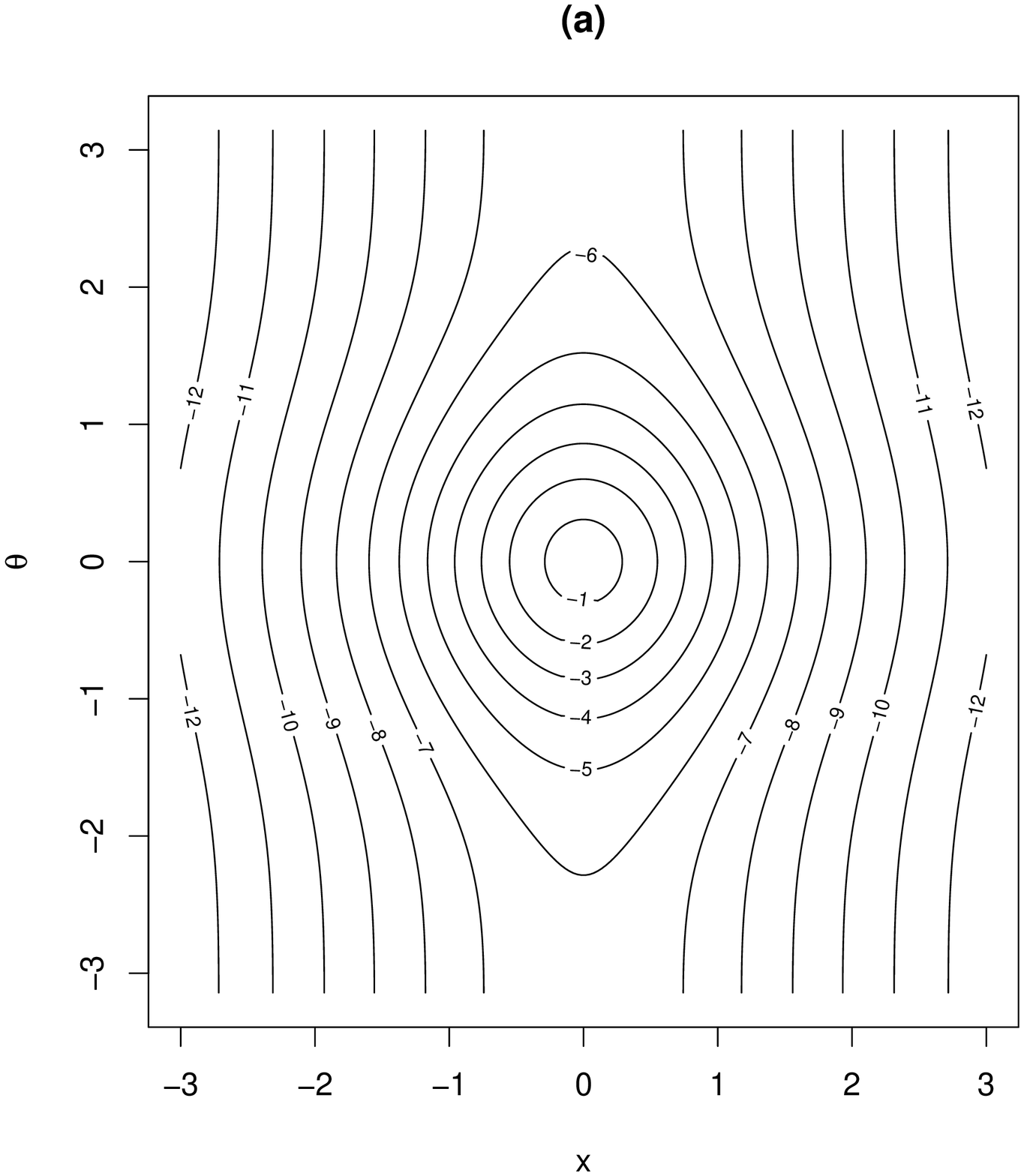}
\includegraphics[width=5cm,clip]{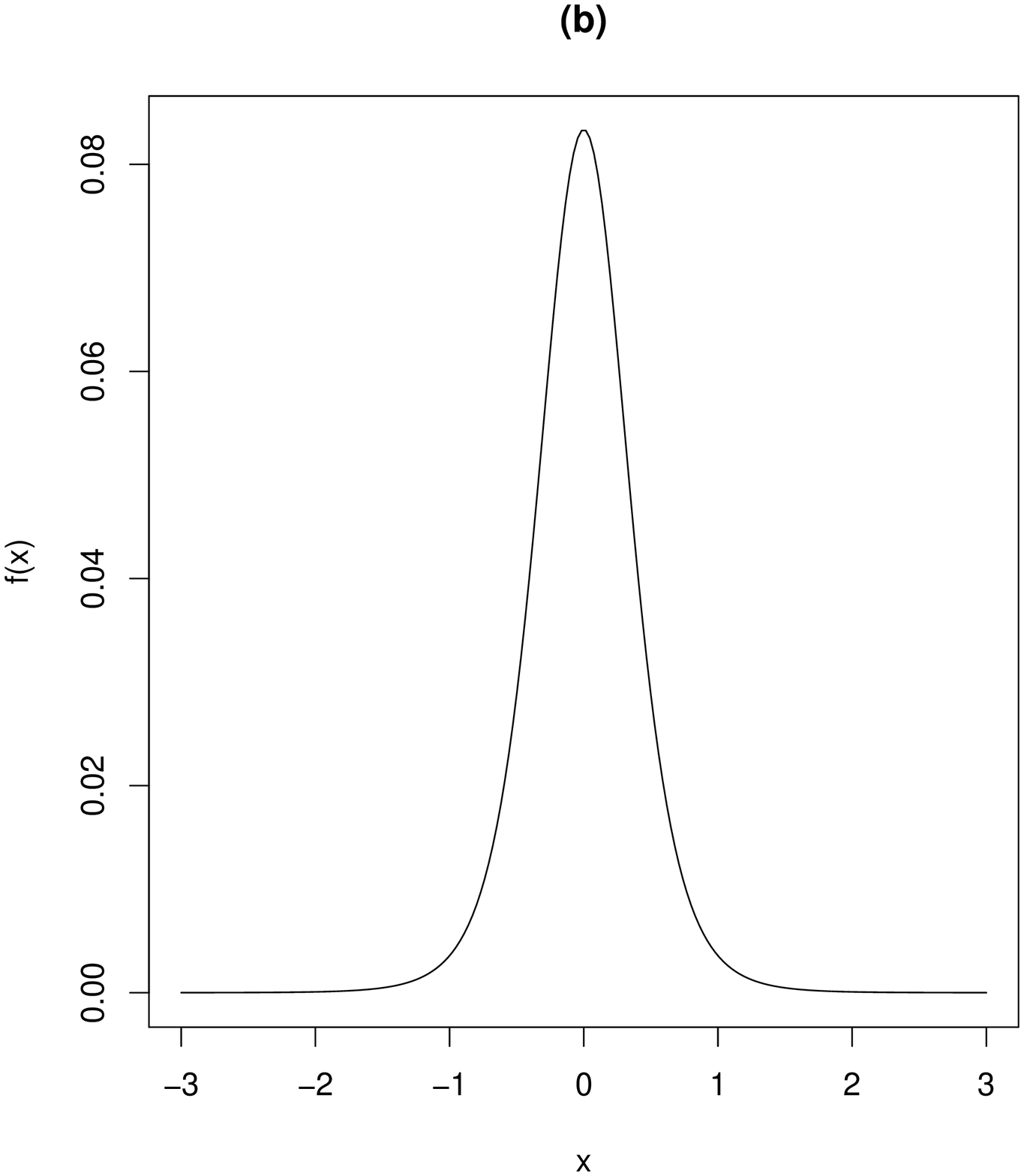}
\includegraphics[width=5cm,clip]{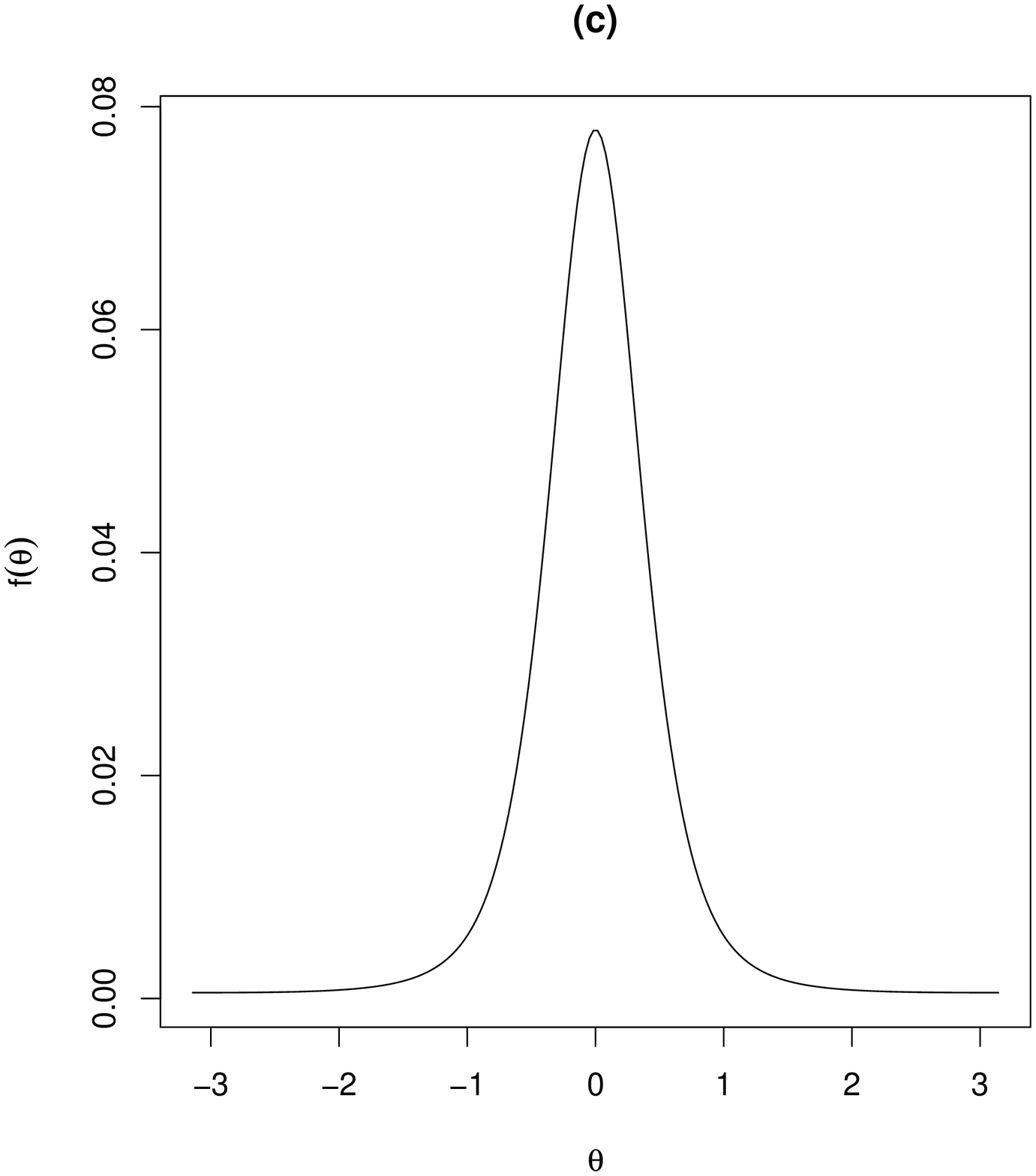}\\
(II)\\
\includegraphics[width=5cm,clip]{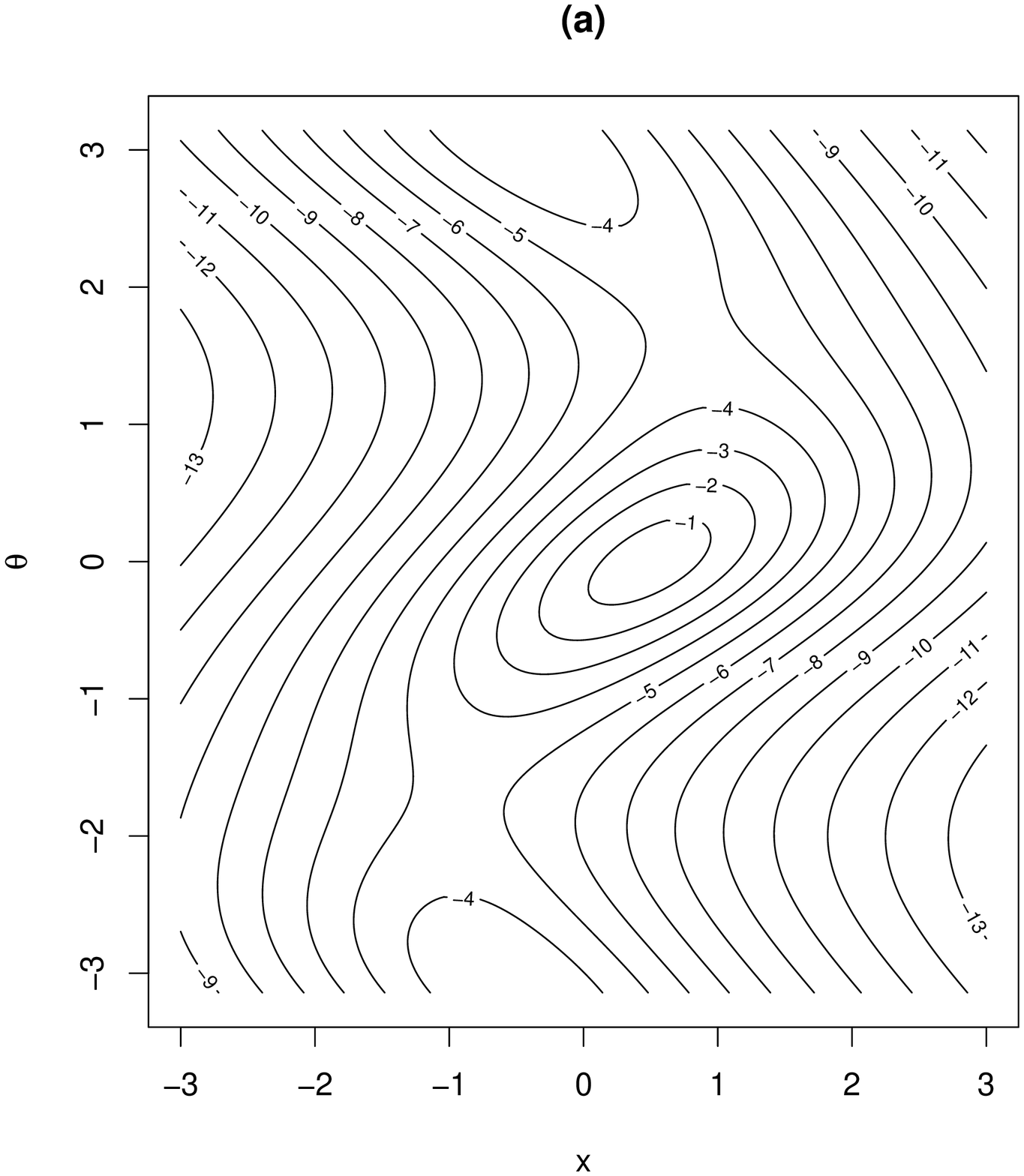}
\includegraphics[width=5cm,clip]{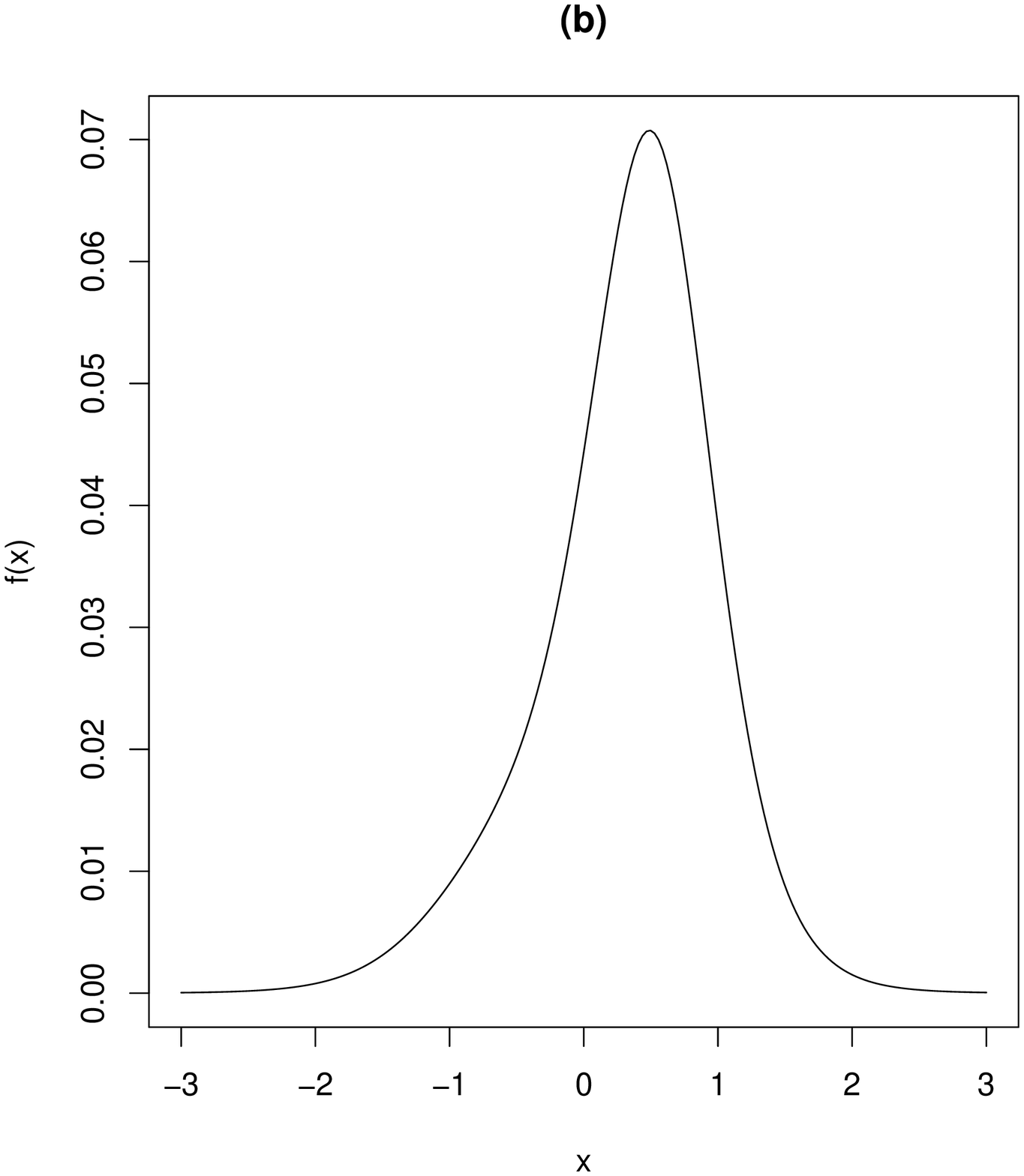}
\includegraphics[width=5cm,clip]{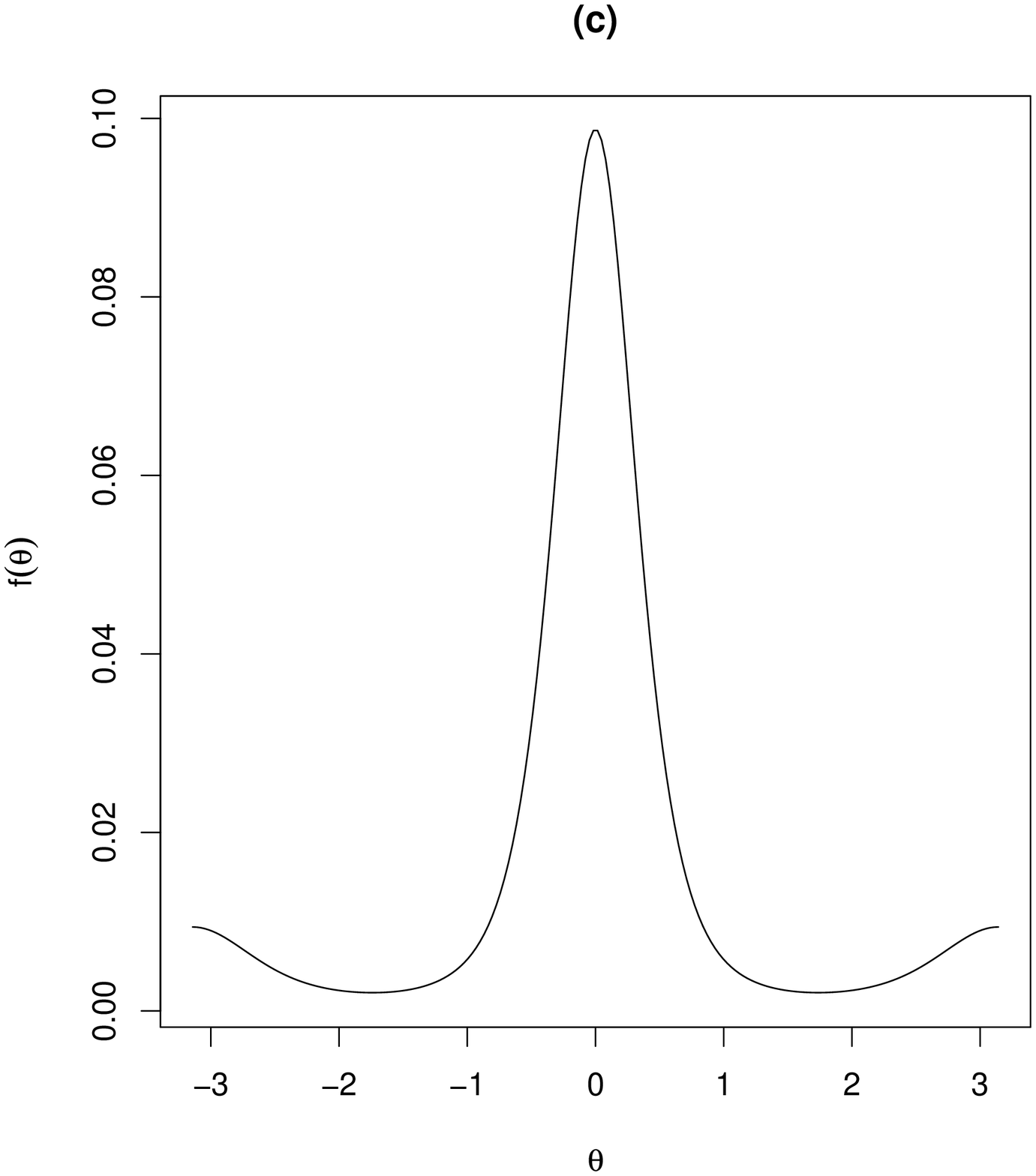}
\caption{(a) Contour plot of log-density, and marginal density plots of (b) the linear random variable $X$ and (c) the circular random variable $\Theta$ for the proposed distribution for: (I) $\kappa_1=0.5$, $\kappa_2=0.1$, $\lambda=0$, and (II) $\kappa_1=0.2$, $\kappa_2=0.3$, $\lambda=1$. 
The other parameters are set as $\sigma=1$,  $\mu=0$, $\nu=\pi/3$, $\mu_1=\mu_2=0$, $\alpha=6$. }
\end{figure}

\subsection{Moments and circular-linear correlation}
We consider moments of the proposed distribution. Let $\alpha_{m,k}=E(\cos m\theta)$ and $\beta_{m,k}=E(\sin k\theta)$ be the trigonometric moments of (\ref{marginal}) under replacement $\alpha/2$ with $\alpha/2-k$, which are obtainable using the results by Siew et al. (2008). Moments of a random vector $(X,\Theta)'$ having (\ref{density}) are given by
\begin{equation}
E[\{X-\mu(\Theta)\}^{2k}\cos m\Theta]=C_{\Theta,k}C^{-1}2^{k+1/2}\sigma^{2k+1}B\left(k+\frac12,\frac{\alpha}{2}-k+1\right)\alpha_{m,k}   \label{gmome1}
\end{equation}
and
\begin{equation}
E[\{X-\mu(\Theta)\}^{2k}\sin m\Theta]=C_{\Theta,k}C^{-1}2^{k+1/2}\sigma^{2k+1}B\left(k+\frac12,\frac{\alpha}{2}-k+1\right)\beta_{m,k},   \label{gmome2}
\end{equation}
where $C_{\Theta,k}$ is obtained by replacing $\alpha$ with $\alpha-2k$ in $C_{\Theta}$ in (\ref{marNC}).

Moreover, we derive another type of moments of a random vector $(X,\Theta)'$ having (\ref{SC2}) given by putting $\kappa_2=0$ in (\ref{density}). After some calculations, the moments for nonnegative integers $k\ (<\alpha/2+1)$ and $m$ turn out to be
\begin{align}
&E[\left\{X-\mu(\Theta)\right\}^{2k}\cos m(\Theta-\mu_1)]\notag \\  
&=\frac{2^{k}\sigma^{2k}B\left(k+\frac12,\frac{\alpha}{2}-k+1\right) (-1)^{m+\frac{\alpha}{2}-k+1}P_{\alpha/2-k}^m\left(-(1-\kappa_1^2)^{-1/2}\right)\left(1-\kappa_1^2\right)^{-(\alpha-2k+2)/4}}{B\left(1/2,\alpha/2+1\right) {_2F_1}\left(\alpha/4+\frac12,\alpha/4+1,1;\kappa_1^2\right)(\alpha/2-k-m+1)_m}, \label{mome}
\end{align}
where $P$ denotes the associated Legendre function (Gradshteyn and Ryzhik, 2007, 8.711.2) defined by
$$
P_{\nu}^m(z)=\frac{(\nu+1)(\nu+2)\cdots(\nu+m)}{\pi}\int_0^{\pi}\left(z+\sqrt{z^2-1}\cos\varphi\right)^{\nu}\cos m\varphi \ {\rm d}\varphi.
$$

Next, we study the circular-linear correlation $R_{x\theta}$ between $X$ and $\Theta$ (cf. Mardia and Jupp, 1999, p. 245) which is defined as
$$
R_{x\theta}^2=\frac{r_{xs}^2+r_{xc}^2-2r_{cs}r_{xs}r_{xc}}{1-r_{cs}^2},
$$
where $r_{xs}={\rm Corr}(X,\cos\Theta),r_{xc}={\rm Corr}(X,\sin\Theta)$ and $r_{cs}={\rm Corr}(\cos\Theta,\sin\Theta)$ are Pearson's correlation coefficients. We only consider the case when $(X,\Theta)'$ has density (\ref{SC2}) for simplicity because the circular-linear correlation of $(X,\Theta)$ with density (\ref{density}) is not feasible to compute analytically.  
Note that the $R_{x\theta}$ in case of (\ref{density}) can be obtained by numerical calculation.
A straightforward calculation shows that 
\begin{equation}
R_{x\theta}^2=\frac{\lambda^2 U}{q+\lambda^2 U},   \label{corr}
\end{equation}
where 
$$
U=\frac{1}{2}\left(1-p_2\right)\sin^2(\mu_1-\nu)+\left\{\frac{1}{2}\left(1+p_2\right)-p_1^2\right\}\cos^2(\mu_1-\nu),
$$
and $p_1,p_2$ and $q$ denote $p_m=E\{\cos m(\Theta-\mu_1)\}, m=1,2,$ and $q=E[\{X-\mu(\Theta)\}^2]$, calculable from (\ref{mome}). 
Note that $U>0$ since $1-p_2>0$ and $(1+p_2)/2-p_1^2={\rm Var}\{\cos(\Theta-\mu_1)\}>0$. We can observe that $R_{x\theta}^2$ is an increasing function of $\lambda$, and $R_{x\theta}^2=0$ if and only if $\lambda=0$. Furthermore, letting $\gamma=\alpha/2\to\infty$, it is seen that (\ref{corr}) reduces to the circular-linear correlation of the distribution proposed by Mardia and Sutton (1978) because (\ref{SC2}) goes to the Mardia--Sutton model. This is confirmed from the fact that $q\to \tau^2$ and $p_m\to I_m(\kappa^{\ast})/I_0(\kappa^{\ast}), m=1,2$ as $\gamma=\alpha/2\to\infty$.

\subsection{Skewness of marginal distribution on the real line}

For the proposed density (\ref{density}), we derive the skewness of the marginal density of $X$ defined as
$$
\gamma_1=\frac{E\left[\left\{X-E(X)\right\}^3\right]}{\left(E\left[\left\{X-E(X)\right\}^2\right]\right)^{3/2}}.
$$
Straightforward calculation shows that
\begin{equation}
\label{sk}
\gamma_1=\frac{3v_2\lambda+v_3\lambda^3}{(v_1\lambda^2+q)^{3/2}},
\end{equation}
where $v_1={\rm Var}\{\cos(\Theta-\nu)\}, \  v_2={\rm Cov}[\{X-\mu(\Theta)\}^2,\cos(\Theta-\nu)]$ and $v_3=E([\cos(\Theta-\nu)-E\{\cos(\Theta-\nu)\}]^3)$, which are calculable from (\ref{gmome1}) and (\ref{gmome2}). We can easily obtain from (\ref{sk}) that $\gamma_1\to v_3/v_1^{3/2}$ as $\lambda\to\infty$ and $\gamma_1=0$ when $\lambda=0$. Figure 4 shows that a graph of skewness as a function of $\lambda$. We see from Figure 4 that the marginal distribution of X of the proposed model can be left and right skewed according to the values of the parameters.

\begin{figure}[!thb]
\centering
\includegraphics[width=7.5cm,clip]{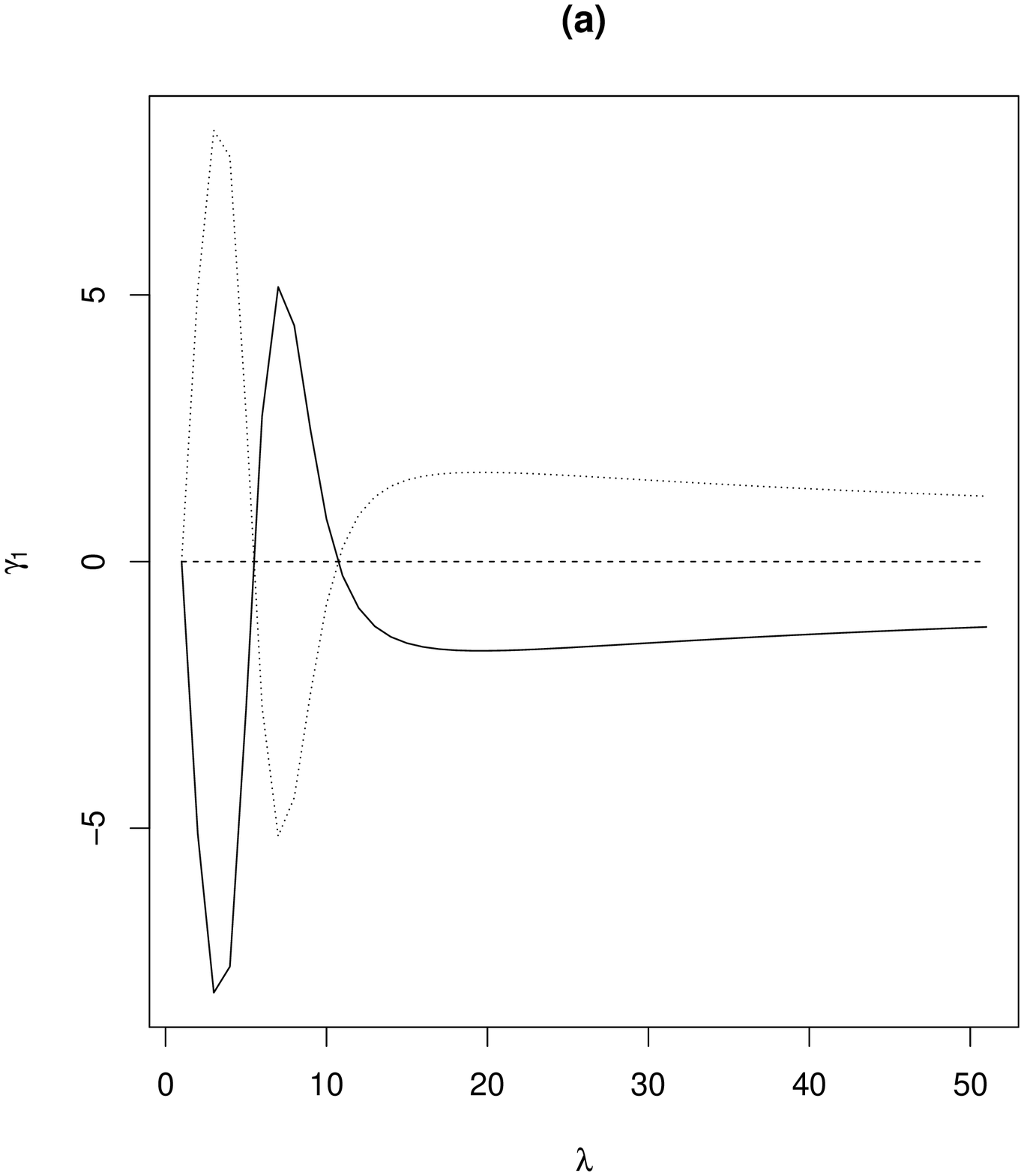}
\includegraphics[width=7.5cm,clip]{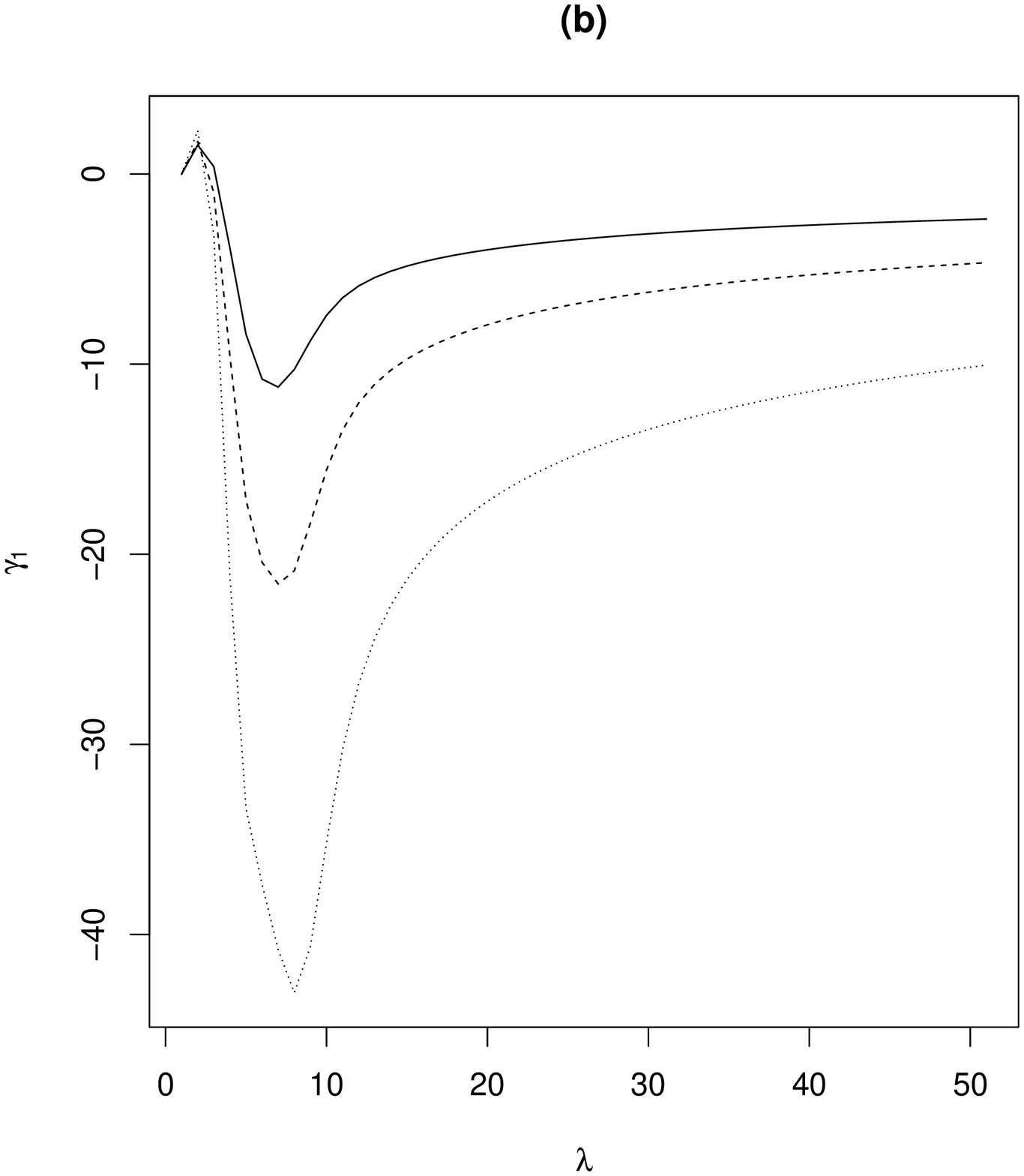} 
\caption{Skewness of the marginal distribution of X as a function of $\lambda$ for (a)  $\sigma=3, \ \mu=0, \ \kappa_1=0.5, \ \kappa_2=0, \ \mu_1=\mu_2=0, \ \alpha=9$ with
$\nu=\pi$ (solid), $\nu=\pi/2$ (dashed) and $\nu=0$ (dotted), and (b) $\sigma=3, \ \mu=0, \ \nu=0, \ \kappa_1=0.3, \  \mu_1=\mu_2=0, \ \alpha=9$ with $\kappa_2=0$ (solid), $\kappa_2=0.1$ (dashed) and $\kappa_2=0.2$ (dotted).}
\end{figure}

\subsection{Regression}

We can derive a circular-linear regression model from the conditional distribution with density (\ref{cond}). 
In fact, the conditional mean of $X$ given $\Theta=\theta$ is
$$
E(X|\Theta=\theta)=\mu(\theta)=\mu+\lambda\cos(\theta-\nu)
$$
and the conditional variance of $X$ given $\Theta=\theta$ is
$$
{\rm Var}(X|\Theta=\theta)=\frac{2\sigma^2}{\alpha}\left\{1-\kappa_1\cos(\theta-\mu_1)-\kappa_2\cos2(\theta-\mu_2)\right\}.
$$
Note that the conditional variance is dependent on $\theta$, i.e. the regression model is possibly heterogeneous. 
As a reduced model, if we let $\gamma=\alpha\to\infty$, we have ${\rm Var}(X|\Theta=\theta)=\tau^2$,
which is independent of $\theta$. Moreover if $\kappa_1=0$ and $\kappa_2=0$ in (\ref{cond}), we have ${\rm Var}(X|\Theta=\theta)=2\sigma^2/\alpha$ and, in this case, we obtain a regression model
$$
x_i=\mu+\lambda\cos(\theta_i-\nu)+\varepsilon_i,  \ \ \ \ \ i=1,\ldots,n,
$$
with random errors $\varepsilon_i$ which are independent and identically distributed according to the generalized $t$-distribution.

\subsection{Maximizing $\beta$-entropy}
The related distribution proposed by Mardia and Sutton (1978) and Kato and Shimizu (2008) can be characterized as the maximum entropy distribution under certain moment conditions. 
{Also a maximum entropy distribution under certain moment conditions relates to the proposed distribution with density (\ref{density}).} We consider the $\beta$-entropy (see Eguchi, 2009, Section 13.2.4) defined as
$$
E(f)=\left\{\int f^{\beta+1}(x,\theta){\rm d}x{\rm d}\theta-\beta-1\right\}\bigg/ \beta(\beta+1).
$$
Then the maximum entropy distribution {subject to constraints on} the moments
$$
E(X^2), \ E(X\cos\Theta), \ E(X\sin\Theta), \ E(\cos p\Theta), \ E(\sin p\Theta), \ p=1,2,
$$
is the distribution with density
{
\begin{equation}
\label{eq:MaxBeta-entropyDensity}
f(x,\theta)\propto \left(1+\beta\left[-\frac{\left\{x-\mu(\theta)\right\}^2}{2\tau^2}+\kappa_1^{\ast}\cos(\theta-\mu_1)+\kappa_2^{\ast}\cos2(\theta-\mu_2)\right]\right)^{1/\beta}.
\end{equation}
If we take $\beta=-2/(\alpha+3)$, (\ref{eq:MaxBeta-entropyDensity}) gives a density related to (\ref{density}).
}

\subsection{Parameter estimation}
We provide a method for calculating the maximum likelihood estimates of the parameters in the generalized $t$-distribution with density (\ref{density}).
When we observe $(x_i,\theta_i), \ i=1,\ldots,n$, the log-likelihood function is given by
\begin{align*}
L(\bpsi)&=-n\log B\left(\frac12,\frac{\alpha}{2}+1\right)-n\log C_{\theta}-\frac n2\log 2-n\log\sigma\\
&-\frac12\left(\alpha+3\right)\sum_{i=1}^n\log\left[1+\frac1{2\sigma^2}\left\{x_i-\mu-\lambda\cos(\theta_i-\nu)\right\}^2
-\kappa_1\cos(\theta_i-\mu_1)-\kappa_2\cos 2(\theta_i-\mu_2)\right],
\end{align*}
where $\bpsi=(\sigma,\mu,\la,\nu,\kappa_1,\mu_1,\kappa_2,\mu_2,\alpha)'$. For obtaining the maximizer of $L(\bpsi)$, we propose the conditional maximization algorithm. We first divide the parameter $\bpsi$ into $\bpsi=(\sigma,\bpsi_1',\bpsi_2')'$, where $\bpsi_1=(\mu,\la,\nu)'$ and $\bpsi_2=(\kappa_1,\mu_1,\kappa_2,\mu_2,\alpha)'$. Given the value of $\sigma$ and $\bpsi_2$, maximizing $L(\bpsi)$ is equivalent to maximizing 
$$
-\sum_{i=1}^n\log\left[C_i+\frac1{2\sigma^2}\left\{x_i-\mu-\lambda\cos(\theta_i-\nu)\right\}^2
\right],
$$
with respect to $\bpsi_1$, where $C_i=1-\kappa_1\cos(\theta_i-\mu_1)-\kappa_2\cos 2(\theta_i-\mu_2)$. Let $\x=(x_1,\ldots,x_n)'$, $\T=(\t_1',\ldots,\t_n')'$ for $\t_i=(1,\cos\theta_i,\sin\theta_i)'$, $\bbe=(\mu,\la\cos\nu,\la\sin\nu)'$ and $\W={\rm diag}(w_1,\ldots,w_n)$ for
$$
w_i=\frac{\sigma^2}{\sigma^2+\left\{x_i-\mu-\lambda\cos(\theta_i-\nu)\right\}^2/2}.
$$
Using the theory of weighted regression (see Andrews, 1974), the maximizer of $\bpsi_1$ given $\sigma$ and $\bpsi_2$ can be obtained as
\begin{equation}\label{psi1}
\widehat{\bbe}=(\T'\W\T)^{-1}\T'\W\x,
\end{equation}
which deduces the maximizer $\bpsih_1$.
Since $\W$ depends on $\bpsi_1$, we calculate $\W$ based on the current values in each iteration. 

Given $\bpsi_1$ and $\bpsi_2$, maximizing $L(\bpsi)$ with respect to $\sigma^2$ is equivalent to solving the following equation:
 \begin{equation}\label{sig}
n\left(\alpha+3\right)^{-1}=\sum_{i=1}^n\frac{\left\{x_i-\mu-\lambda\cos(\theta_i-\nu)\right\}^2}{\left\{x_i-\mu-\lambda\cos(\theta_i-\nu)\right\}^2+C_i\sigma^2},
\end{equation}
which deduces the maximizer $\hat{\sigma}$.

Finally for maximizing $\bpsi_2$ under given $\sigma$ and $\bpsi_1$, we maximize 
\begin{equation}\label{psi2}
-n\log B\left(\frac12,\frac{\alpha}{2}+1\right)-n\log C_{\theta}-\frac12\left(\alpha+3\right)\sum_{i=1}^n\log\left\{D_i-\kappa_1\cos(\theta_i-\mu_1)-\kappa_2\cos 2(\theta_i-\mu_2)\right\}
\end{equation}
with respect to $\bpsi_2$, where $D_i=1+(2\sigma^2)^{-1}\left\{x_i-\mu-\lambda\cos(\theta_i-\nu)\right\}^2$. This maximization problem is quite similar to obtaining the maximum likelihood estimates of the generalized $t$-distribution on the circle (Siew et al., 2008), so that we can obtain the maximizer $\bpsi_2$ given $\sigma$ and $\bpsi_1$.
Note that we carried out the maximization of (\ref{psi2}) with use of numerical integration for getting the value of $C_{\theta}$.

Therefore, the proposed estimation method is described in the following.

\ \\
{\bf Estimation Algorithm}
\begin{itemize}
\item[1.]\ \ Determine the initial values $\bpsi^{(0)}$. Set $k=0$.
\item[2.]\ \ Compute  $\W$ based on $\sigma^{(k)}$ and $\bpsi_1^{(k)}$, then obtain $\bpsi_1^{(k+1)}$ based on (\ref{psi1}).
\item[3.]\ \ Compute $\sigma^{(k+1)}$ by solving (\ref{sig}) with $\bpsi_1=\bpsi_1^{(k+1)}$ and $\bpsi_2=\bpsi_2^{(k)}$.
\item[4.]\ \ Compute $\bpsi_2^{(k+1)}$ by maximizing (\ref{psi2}) with $\bpsi_1=\bpsi_1^{(k+1)}$ and $\sigma=\sigma^{(k+1)}$.
\item[5.]\ \ Set $k=k+1$ and go to Step 2 (until numerical convergence).
\end{itemize}

\section{Empirical application}

\begin{figure}[!thb]
\begin{center}
\includegraphics[width=8cm,clip]{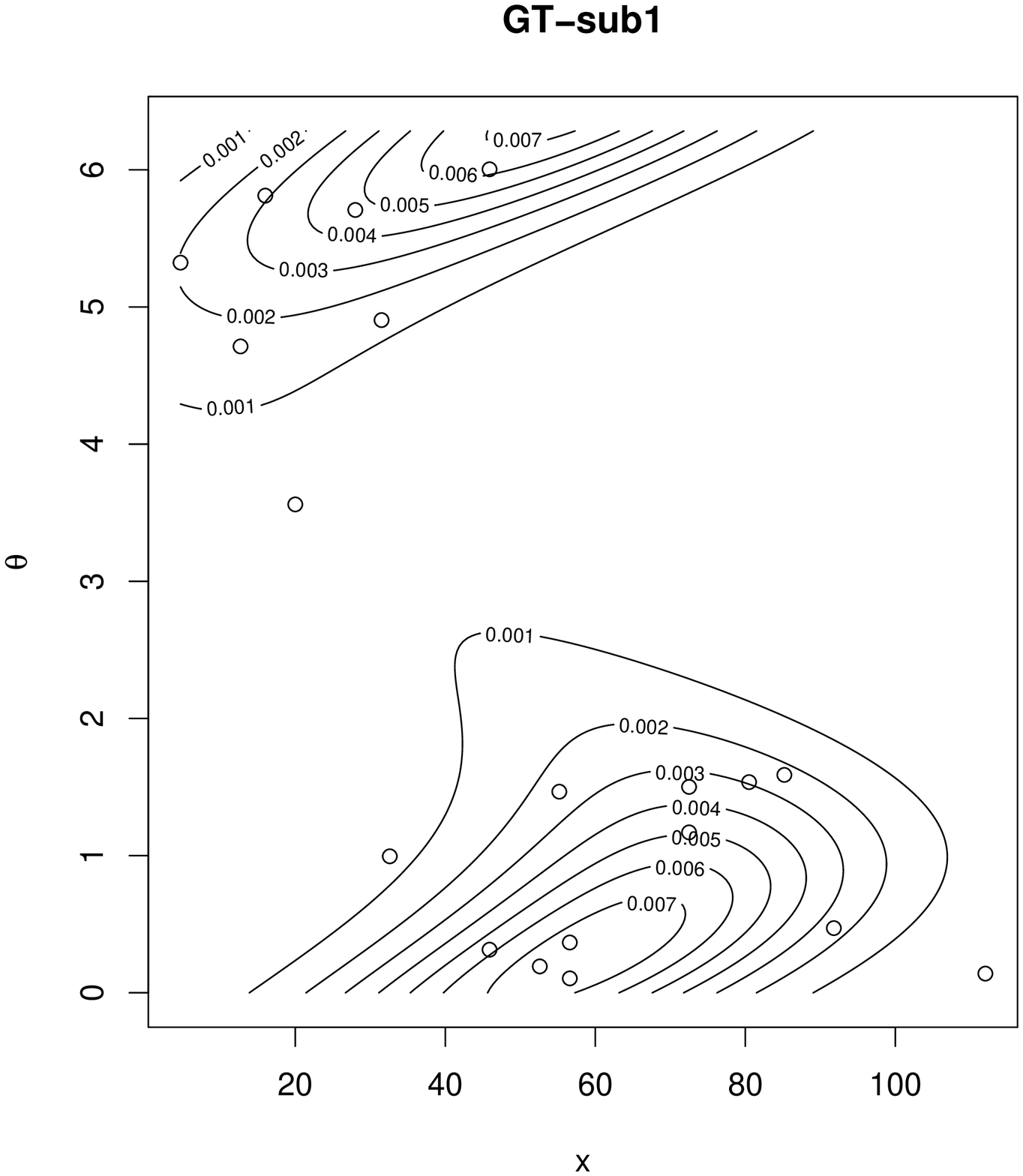}
\includegraphics[width=8cm,clip]{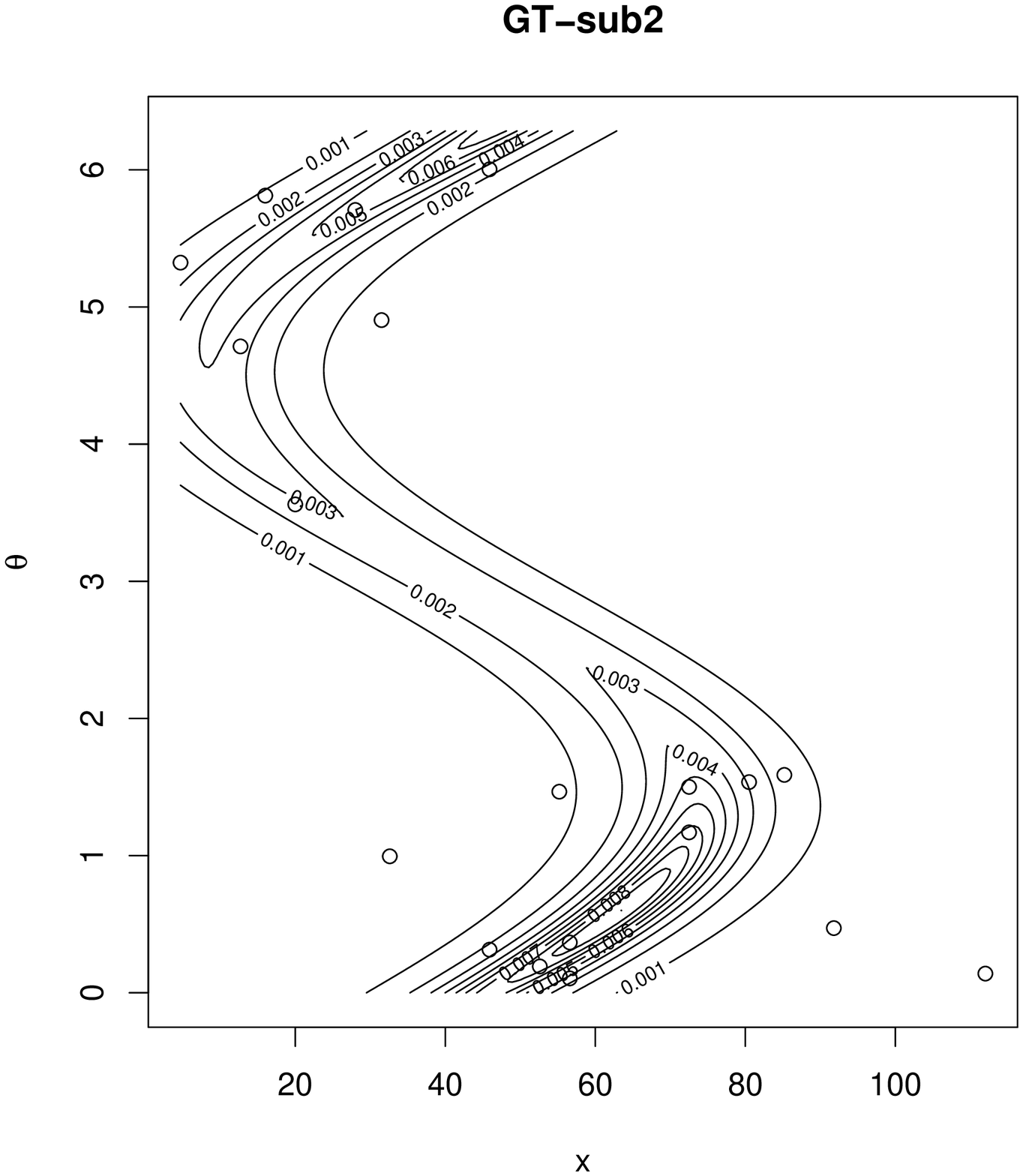}
\end{center}
\caption{Sample plot and contour plot of the fitted density function.}
\label{fitted}
\end{figure}

For an illustrative example, we consider a cylindrical dataset given in Johnson and Wehrly (1977) on the wind direction and ozone level taken at 6:00 pm at four-day intervals between April 18th and June 29th, 1975 at a weather station in Milwaukee with $19$ samples.
We fitted the proposed generalized $t$-distribution and its submodels.
For submodels of (\ref{density}), we consider two cases, namely, $\kappa_2=0$ (GT-sub1) given in (\ref{SC2}) and $\mu_2=\mu_1+\pi/4$ with $2|\kappa_2|<\kappa_1$ (GT-sub2) discussed in the end of Section 3.3.
When we carry on the estimation algorithm given in Section 3.8, we repeat the algorithm until the difference between update and current values are smaller than $0.001$.
For comparison, we also fitted the member of the exponential family given by Kato and Shimizu (2008) with density
$$
f(x,\theta)\propto \exp\left\{-\frac{(x-\mu(\theta))^2}{2\sigma^2}+\kappa_1\cos(\theta-\mu_1)+\kappa_2\cos2(\theta-\mu_2)\right\},
$$
which is reparametrized form of (\ref{KS-dist}) with $\kappa_1^{\ast}=\kappa_1$, $\kappa_2^{\ast}=\kappa_2$ and $\tau^2=\sigma^2$.
Table \ref{ML} provides the maximum likelihood estimates of the parameters, AIC values, and the multivariate Kolmogorov-Smirnov statistics of goodness of fit testing given by Justel et al. (1997).
In Table 1 of Justel et al. (1997), the bivariate Kolmogorov-Smirnov statistics distributions are reported.
From the table, the upper $5\%$, $10\%$ and $25\%$ percentiles are $0.362$, $0.335$ and $0.292$, respectively, for $20$ sample cases.
Thus the percentiles in 19 sample cases are smaller than these values, so that the fitted four models are not rejected with $5\%$ significance level.
Judging from AIC, we see that the two submodels of the proposed distribution, GT-sub1 and GT-sub2, gives better fits than the Kato and Shimizu distribution. 
Figure \ref{fitted} shows scatter plots of the data and contour plots of the fitted densities of two submodels.

\begin{table}[!thb]
\caption{Maximum likelihood estimates of the parameters, the maximum log-likelihood (MLL), and AIC values of model (\ref{density}) (GT) and its submodels and the Kato and Shimizu model (\ref{KS-dist}) (KS) fitted to the data from Johnson and Wehrly (1977).}
\begin{center}
\begin{tabular}{cccccccccccccc}\toprule
Model     & $\mu$ & $\la$ & $\nu$ & $\sigma$ & $\kappa_1$ & $\mu_1$ & $\kappa_2$ & $\mu_2$ & $\alpha$ & AIC & g.o.f\\
\midrule
GT & 41.38 & 31.14  & 1.25 & 68.90 &  0.11 &  6.28 &  0.03  & 1.35&  23.84 & 417.79 & 0.180 \\
GT-sub1&41.37& 31.17 &  1.24 & 74.81 &  0.08&   0.30 & --- & --- & 28.28 &240.20 & 0.314\\
GT-sub2&41.01 & 32.70 &  1.41 &  6.08 &  0.49 &  0.19 &  0.15 &---& -1.00  &203.34 & 0.359\\
KS & 41.24 & 31.38 &  1.20 & 19.77 &  1.41 &  0.19  & 0.35 &  1.45 & ---  & 241.42 & 0.282\\
\bottomrule
\end{tabular}
\label{ML}
\end{center}
\end{table}

\section{Conclusions}
In this paper, we derived the distributions on the cylinder based on a trivariate $t$-distribution.
The derived distribution is considered as a cylindrical extension of the generalized $t$-distribution on the circle proposed by Siew et al. (2008) and includes the exponential family given by Kato and Shimizu (2008).
We investigated some properties of the proposed distribution including the marginal and conditional distributions, circular-linear correlation and the algorithm for parameter estimation.
We applied our proposed distribution to the data set of the wind direction and ozone level given in Johnson and Wehrly (1977), and we confirmed that the proposed distribution gave a better fit than the distribution given by Kato and Shimizu (2008).

\vspace{0.5cm}
\noindent
{\bf Acknowledgement}

We would like to thank the two reviewers for many valuable comments and helpful suggestions which led to an improved version of this paper. 
The first author was supported in part by Grant-in-Aid for Scientific Research (10076) from Japan Society for the Promotion of Science (JSPS).
The work of the third author was supported by JSPS KAKENHI Grant Number 25400218.


\end{document}